\documentclass[apj]{emulateapj}
\usepackage{natbib}
\shorttitle{Assembly of M81 and the Local Group}
\shortauthors{Nichols \&{} Bland-Hawthorn}

\begin{document}

\title{The Epoch of Assembly of Two Galaxy Groups: A comparative study}
\author{Matthew Nichols\altaffilmark{1}}
\email{matthew.nichols@epfl.ch}
\affil{Laboratoire d'Astrophysique, \'Ecole Polytechnique F\'ed\'erale de Lausanne (EPFL), Observatoire de Sauverny, 1290 Versoix, Switzerland}
\altaffiltext{1}{Sydney Institute for Astronomy, School of Physics, The University of Sydney, NSW 2006, Australia}
\and
\author{Joss Bland-Hawthorn}
\affil{Sydney Institute for Astronomy, School of Physics, The University of Sydney, NSW 2006, Australia}

\begin{abstract}
Nearby galaxy groups of comparable mass to the Local Group show global variations that reflect differences in their evolutionary history.
Satellite galaxies in groups have higher levels of gas deficiency as the distance to their host decreases.
The well established gas deficiency profile of the Local Group reflects an epoch of assembly starting at $z\lesssim 10$.
We investigate whether this gas deficiency profile can be used to determine the epoch of assembly for other nearby groups.
We choose the M81 group as this has the most complete inventory, both in terms of membership and multi wavelength observations.
We expand our earlier evolutionary model of satellite dwarf galaxies to not only confirm this result for the Local Group but show that the more gas-rich M81 group is likely to have assembled at a later time ($z\lesssim1-3$).
\end{abstract}

\keywords{galaxies: dwarf --- galaxies: interactions --- galaxies: individual (M81) --- Galaxy: evolution --- Galaxy: halo --- Local Group}
\section{Introduction}

The predominant cosmological model of structure formation---dark energy and cold dark matter, $\Lambda$CDM---dictates that galaxy sized objects build up hierarchically through the merger of smaller structures.
Throughout this build-up, these small structures are either destroyed, incorporated into the galaxy's halo, or become satellite subhalos of the host galaxy \citep{Moore1999}.
The destroyed systems and those which survived to become satellites were likely similar, with the satellite subhalos becoming the modern day dwarf galaxies \citep{Frebel2010}.

That dwarf galaxies are likely relics of early galaxy formation has made them prominent targets for observational and theoretical studies of galactic formation. These include analysis of star formation thresholds \citep{deBlok2006,Ekta2008,Kuhlen2012}, how the earliest galaxies formed \citep{Ricotti2008,Tolstoy2011,Frebel2012}, the stellar yields of the first stars \citep{Karlsson2012,Karlsson2013}, the strength of reionization \citep{Barkana1999,Somerville2002,Hoeft2006,Sawala2012,Lunnan2012}, and distinguishing between different forms of dark matter \citep{Bode2001,Maccio2010,Lovell2012}. Satellite dwarfs may also be a guide to the accretion history of their host galaxy.
With cosmological models showing large amounts of variation in the mass accretion history of galaxies \citep{McBride2009,Boylan-Kolchin2010}, variations in the redshifts at which satellite galaxies first fell in would be unsurprising.

Determining the accretion time of dwarf galaxies has involved either (i) calculating the orbits of dwarfs backwards in time \citep[e.g.][]{Besla2007,Lux2010,Bekki2011} or (ii)  statistical methods comparing infall epochs and the present day orbital properties of dwarf galaxies \citep{Rocha2012,Teyssier2012}. 
The accretion history of the Local Group can be derived by comparing the observed distance-H{\small I} deficiency relation in satellite galaxies \citep{Grcevich2009} with evolution models.
A recent study has found that the Milky Way and M31 have been accreting satellites since $z\lesssim 10$ \citep[hereafter NBH]{Nichols2011}.

Such a relation between the distance to the host and fraction of dwarfs that are deficient in H{\small I} is not unique to the Local Group, with other nearby groups showing a range of related trends \citep{Karachentsev2007,Rasmussen2012,Bahe2013,Crnojevic2012,Roychowdhury2012,Chiboucas2013}.
We explore these trends to compare the likely differences in accretion history between these nearby groups.
In order to achieve this, we use the model presented in NBH, but with an updated treatment of dwarf stripping to allow for the presence of disks and a larger subhalo mass range.

We use this new model to examine the accretion history of nearby galaxy groups based upon this distance/gas-deficiency relation to examine the relative mass accretion histories between the Local Group and the M81 group another nearby group. In \S 2, we present our improved dynamical model for an evolving galaxy group. In \S 3, we present the results of comparative study between the Local Group and the M81 group, and we present our conclusions in \S 4.

\section{Model}
To determine the final state of dwarf galaxies, a toy model of a dwarf galaxy is evolved on a random orbit from a redshift of $z=10$.
Here we describe the orbital model briefly before discussing the initial conditions of the dwarf, the heating and cooling of the gas and the stripping of the gas.
Throughout the model we track $8$ variables: four describing the galactocentric distance, angle of the dwarf and their derivatives, ($r$, $\theta$, $\dot{r}$, $\dot{\theta}$); $R_{\rm d}$, the radial size of the dwarfs disk; $M_{\rm d}$ a parameter which describes the mass distribution of the dwarfs disk; $R_{\rm h}$, the radial extent of the dwarfs hot halo and; $X_{\rm c}$, the ratio of the cold gas mass in the disk to the total disk mass.

\subsection{Orbital Model}
For each run the dwarf galaxy has a random orbit determined by a phase, perigalacticon and circularity.
The perigalacticon and circularity are selected by sampling the $z=0$ infall distributions of \citet{Wetzel2011} in an identical fashion to \citet{Nichols2011b}.
We restrict the perigalacticon to a minimum value of $10$~kpc.
Dwarfs that end up closer in will have a much larger contribution from the disk of the Galaxy and will likely have been tidally destroyed.
The distributions are assumed to continue past the virial radius to a maximum value of eight times the virial radius ($\sim$$2$~Mpc).
Given the exponential nature of the distribution, this limit is likely to have minimal impact and is chosen only to avoid calculation of orbits wholly beyond the radius considered in the comparison.
Dwarfs beyond the virial radius will never be considered to have infallen but do add a minor contribution to the fraction of gas deficient galaxies beyond the virial radius.
The infall distribution with redshift is not constant.
However, as the orbits would invariably change as the host halo grows (even ignoring dynamical friction) we assume the $z=0$ distribution is representative of the final distribution of orbits.

A random phase is chosen by evolving the orbit from perigalacticon for $10$~Gyr (an arbitrary amount of time) inside a static Einasto halo of mass equal to the model host galaxy at $z=0$.
A final position is then chosen by selecting the position at a random point in time up to the time of first apogalacticon (or $10$~Gyr, whichever is later).
Such a method accounts for the non-elliptical nature of orbits in an Einasto halo, while only missing a small fraction of the parameter space (the portion of the orbits that exceeds the age of the Universe).
The dwarf is then randomly chosen to be heading towards or away from the host galaxy at $z=0$.

The initial position (at $z=10$) is then determined by evolving the final position backwards inside a growing Einasto potential where the equation of motions are given in NBH.
The mass of the host galaxy is assumed to evolve according to the prescription given in \citet{Boylan-Kolchin2010}
\begin{equation}
M(z) = M_0(1+z)^{2.23}\exp[-4.90(\sqrt{1+z}-1)].
\end{equation}
The profile of the host galaxy is assumed to be an Einasto profile with an Einasto parameter $\alpha=0.16$ \citep[for all mass ranges considered from dwarf galaxy to a few times the Milky Way, the parameter $\alpha$ is dominated by the constant term in fits to {\em N}-body simulations; ][]{Gao2008,Duffy2008}. The concentration of the host halos is given by \citep{Duffy2008}
\begin{equation}
c_{\rm vir} = \frac{8.82}{(1+z)^{0.87}}\left(\frac{M_{\rm vir}}{{2\times10^{12}h~M_\odot}}\right)^{-0.106}.
\end{equation}

The initial position is then used to evolve the orbit forwards in time taking into account the evolution of the dwarf galaxy.

\subsection{Dwarf Galaxy Initial Conditions}
Dwarf galaxies are low mass, low luminosity galaxies with a broad array of morphologies.
Despite the wide range of morphologies, it is likely that at least the majority of satellite dwarfs are examples of different evolutionary states of gas rich, disk containing ``early-type'' dwarfs \citep{Mayer2001,Grebel2003,Pasetto2003,D'Onghia2009,Helmi2012}.
We hence model all dwarfs as if they originated from a constant template: a spherical dark matter halo containing a mixed warm/cold gas disk with a hot gas halo and the entire dwarf embedded inside a host galaxy halo filled with its own hot gas.

The dark matter halo of the dwarf galaxy is chosen to be a static Einasto profile with the mass a free parameter.
An Einasto profile best matches the properties of dark matter halos arising through large {\em N}-body, dark matter simulations \citep{Springel2008}.
All halos are modelled with an Einasto parameter of $\alpha=0.16$ and a concentration at $z=0$ of $c_{\rm dw,0}=35$, the approximate median of dwarf galaxy sized halos in \citet{Colin2004}.
The choice of an Einasto profile is not without problems.
Although the Einasto profile is the best match to dark matter simulations, observational results suggest that the central portion of dwarf galaxy dark matter halos may be constant density although this remains controversial with both ``cored'' and ``cuspy'' profiles seen \citep{Simon2005,DelPopolo2012,Wolf2012}.
Such a profile may arise through the halo consisting of warm dark matter rather than cold dark matter \citep{Bode2001} or through baryonic processes such as non-adiabatic expulsion of gas from supernova \citep{Pontzen2012}.
We do not consider either situation in this paper, but we note that our method could be extended (with another free parameter) in principle to accomodate either case.
The impact of the latter case is difficult to predict, since ram pressure stripping should remove gas reducing the star formation over time (and therefore the amount of supernova) such a process may result in the density profiles of dwarf galaxies being fundamentally different from host galaxy to host galaxy.
This is a possibility not explored in this paper although such processes should result in more ram pressure stripping and therefore a lower final gas mass in all dwarfs.

The disk of the dwarf galaxy is assumed to consist of a cold/warm mixture of gas with an initial hydrogen mass, $M_{\rm gas}$, equal to the baryon (after accounting for He)/dark matter ratio at $z=0$ \citep[$M_{\rm gas}=0.12~M_{\rm dw,0}$;][]{Komatsu2011}.
The disk is chosen to be an exponential disk three scale lengths in size \citep[a reduction in density of $95\%$ and towards the limit that the disk of dwarf galaxies have been seen to continue out to][]{Hunter2011}, with the scale length of the gas assumed to be proportional to the scale length of the optical disk \citep{Bigiel2012}.
The scale length of the disk, $R_{sd}$, is then approximately given by \citep{Bigiel2012}
\begin{equation}
R_{sd} = 0.61\left(\frac{M_{\rm gas}}{172~M_\odot}\right)^{1/2}~{\rm pc}.
\end{equation}
The exponential profile of the disk then gives a surface density 
\begin{equation}
  \Sigma_{\rm gas}(R) = \frac{M_{d}}{2\pi{}R_{sd}^2}\exp(-R/R_{sd}),
\end{equation}
where $M_{d}$ is related to $M_{\rm disk}=M_{\rm gas}+M_{\rm hot}$ through
\begin{equation}
  M_{d} = \frac{M_{\rm disk}}{R_{sd}}\left(R_{sd} -[R_{\rm d}+R_{sd}]\exp[-R_{\rm d}/R_{sd}]\right),
\end{equation}
where $R_{\rm d}$ is the radial size of the disk (initially $R_{\rm d} = 3R_{sd}$).
We also assume that initially there is a negligibly small hot halo of the dwarf and therefore $M_{\rm disk}=M_{\rm gas}$.

The disk is initially assumed to be split equally between a cold gas phase and warm gas phase (therefore, $X_{\rm c}=0.5$ initially) with respective temperatures of $T_{\rm c}=400$~K and $T_{\rm w}=1.2\times10^4$~K.
The central density of cold gas is assumed to be $n_{\rm c}=1.5$~cm$^{-3}$ in line with the observed densities in Leo T \citep{Ryan-Weber2008,Grcevich2009} and the density of warm gas to be $n_{\rm w} = \frac{1}{2}\frac{T_{\rm c}}{T_{\rm w}}\sim0.17$~cm$^{-3}$.

The size of the hot halo is assumed to be initially small (but non-zero for numerical reasons) and is set to be $R_{\rm h} = 0.01$~kpc.
The temperature of any hot halo that forms is assumed to be constant at $1\times10^{6}$~K with a central density of $n_{\rm h}= \frac{1}{2}\frac{T_{\rm c}}{T_{\rm h}}\sim2\times10^{-4}$~cm$^{-3}$. 

\subsection{Heating and Cooling}

The disk of a dwarf galaxy is unlikely to experience a steady balance of heating and cooling.
Being of low mass without densities large enough to guarantee self-shielding such dwarfs will experience a variety of heating processes throughout their lifetime.
The most powerful of these heating processes consist of the host galaxy radiation field, the extragalactic UV field and the dwarfs own internal star formation.

The host galaxy radiation field and the extragalactic UV field are modelled as in NBH, i.e. a host galaxy emitting (after correcting for the escape fraction) $2.6\times10^{51}$~photons~$s^{-1}$ at $13.6$~eV \citep{JBH1999} scaled according to star formation history of the Milky Way given by \citet{Just2010} and an extragalactic radiation field given by \citet{Faucher-Giguere2009}.

The internal star formation in the dwarfs disk is given by the Kennicutt-Schmidt law.
Although the typical threshold of star formation is $M_{\rm{H\scriptsize{I}}}\sim$$10^{21}$~cm$^{-3}$ inside larger galaxies, there is little indication that such a threshold is present inside all dwarf galaxies \citep{Roychowdhury2009}.
We choose a threshold of $N_{\rm HI}=10^{20}$~cm$^{-2}$ for the main model and discuss the impact of varying thresholds in \S\ref{ssec:SFthresh}.

The impact of the star formation and extragalactic UV field is precalculated for a range of disk sizes, redshifts and column densities considering only H and He using the prescription of \citet{Wolfire1995b}.
The results are interpolated from this table inside the model in order to reduce the computational time.
The cooling is also precomputed assuming metal line cooling at a metallicity of $0.1$~$Z_\odot$ with the fits provided by \citet{Schure2009} and He collisional cooling accounted for according to \citet{Dalgarno1972}.
The cooling is assumed to happen equally over the entire disk, with the disk height assumed to be constant across the disk with the density of gas changing radially in line with the surface density.

These processes result only in a change to $X_{\rm c}$ according to
\begin{equation}
  \dot{X}_{\rm c} = \frac{\dot{M}_{\rm cool} - \dot{M}_{\rm heat}}{M_{\rm disk}}.
\end{equation}

In addition to ionizing and heating cold gas the star formation also induces galactic winds from radiation pressure, supernovae and stellar winds.
Inside dwarf galaxies the latter two cases are likely to be dominant \citep{Hopkins2012}.
To model these winds we extend the fitting formula from \citet{Hopkins2012} down to the dwarf galaxy scales considered here, with the magnitude of mass loss given by 
\begin{equation}
\dot{M}_{\rm wind} = 10~{\rm SFR}~V^{-1.1}_{{\rm circ}, 100}(r)\Sigma^{-0.6}_{{\rm HI}, 10}(r)~M_\odot~\rm{yr}^{-1},
\end{equation}
where ${\rm SFR}$ is the star formation rate, $V_{{\rm circ},100}$ is the circular velocity of the dwarf in units of $100$~km~s$^{-1}$ and $\Sigma_{{\rm HI},10}$ is the gas density in units of $10$~M$_\odot$~pc$^{-2}$ calculated at the wind origin.
We assume the wind always launches from the origin and hence may slightly overestimate the magnitude of the wind.

The wind is assumed to lower the height of the disk---that is, takes away mass from the disk while leaving the density of gas constant---while adding to the hot halo.
Much of the wind will in reality be entrained material rather than hot gas.
Such material may fall back to the disk in a galactic fountain, however, it is likely that the majority will be removed via ram pressure stripping before it has a chance to fall back to the disk or it will be destroyed through conduction in the dwarfs hot halo.
Such a wind therefore induces a change in $M_{d}$ and $R_{\rm h}$ of
\begin{eqnarray}
\dot{M}_{d} &=& -\dot{M}_{\rm wind}\frac{R_{sd}}{R_{sd}-(R_{\rm d}+R_{sd})\exp(-R_{\rm d}/R_{sd})},\\
\dot{R}_{{\rm h, !RPS}} &=& \frac{\dot{M}_{\rm wind}}{4\pi{}R_{\rm h}m_{\rm H}n_{\rm h}}f_{\rm gas}(R_{\rm h}/R_{s,dw},v_{s,dw}),
\end{eqnarray} 
where $\dot{R}_{h,{\rm !RPS}}$ is the change in the dwarfs halo radius, not accounting for ram pressure stripping; $f_{\rm gas}(x,v_s)$ is the hydrostatic gas distribution for an Einasto halo \citep[see ][]{Sternberg2002,Nichols2009}; $R_{s,dw}$ is the scale radius of the dwarf's dark matter halo and; $v_{s,dw}$ is the scale velocity of the dwarf's dark matter halo.

\subsection{Ram Pressure Stripping}

As the dwarf is moving relative to the host galaxy hot halo, it will experience a ram pressure force acting upon both its disk and the hot halo.
We model the hot halo of the host galaxy by assuming it follows an exponential profile of form $\rho_{\rm gal} \propto \exp[-V/c_{s}^2]$ where, $V$ is the potential of the host galaxy at the dwarfs point and $c_s$ is the speed of sound inside the hot halo (assumed to be comprised of a primordial mixture of H and He with a temperature of $2\times10^6$~K).
The factor of proportionality is determined by setting a Milky Way sized host galaxy to have a number density of $3\times10^{-4}$~cm$^{-3}$ at a distance of $50$~kpc.
Such a value is consistent with calculations of \citet{Battaglia2005}, \citet{JBH2007} and \citet{Kaufmann2009}.

To calculate ram pressure stripping the hydrostatic halo of the dwarf is approximated as an isothermal sphere, with the ram pressure condition ($S_{\rm RPS}$) for the hot halo then being \citep{McCarthy2008}
\begin{equation}
S_{\rm RPS,h} \rho_{\rm gal}(r)v^2 - \frac{\pi}{2}\frac{GM_{dw}(R_{\rm h})\rho_{\rm halo}(R_{\rm h})}{R_{\rm h}} > 0,
\end{equation}
where $v^2=\dot{r}^2+(r\dot{\theta})^2$, $M_{dw}(R_{\rm h})$ is the mass of the dwarf enclosed within the dwarfs hot halo outer edge,  $\rho_{\rm halo}(R_{\rm h})$ is the density of the dwarfs hot halo at this edge and ram pressure stripping occurs if $S_{\rm h}$ evaluates true.
We ignore any potential hydrodynamic interaction between the disk and the halo which may have a minor protective effect \citep{Bekki2009}.

The disk is assumed to be protected from ram pressure stripping until the dwarf's hot halo is reduced beyond the borders of the disk (i.e. ram pressure stripping only occurs if $R_{\rm d}>R_{\rm h}$ although this condition is nearly always satisfied when the disk would be stripped).
If this is true, the condition for ram pressure stripping is then given by \citep{McCarthy2008}
\begin{equation}
S_{\rm RPS,d} = \rho_{\rm gal}(r)v^2 - g_{\rm max}(R_{\rm d})\Sigma_{\rm disk}(R_{\rm d}) > 0,
\end{equation}
where $g_{\rm max}(R_{\rm d})$ is the maximum restoring force of the gas in the direction of $v$ and $\Sigma$ is the projected gas density.
For simplicity we assume that the disk is moving face on, although there is little difference for a wide range of angles \citep{Quilis2000,Jachym2009}.
For an Einasto halo with $\alpha=0.16$, $g_{\rm max}(R_{\rm d})$ is well fit (to at least a few dwarf scale radius) by
\begin{equation}
g_{\rm max}(R_{\rm d}) = \frac{6.63\times10^{-4}v^2_{s,dw}}{R_{s,dw}}10^{15-1.03(R_{\rm d}/R_{s,dw})^{0.41}}.
\end{equation}
If the disk has expanded past the point of maximum gravitational attraction, $h_{\rm max}$, the value of $g_{\rm max}$ is weighted by calculating the average value of the gravitational attraction (with all mass below $h_{\rm max}$ assumed to have $g_{\rm max}$ attraction).
$h_{\rm max}$ is well fit by
\begin{equation}
h_{\rm max} = \sqrt{2.11\left(\frac{R_{\rm d}}{R_{s,dw}}\right)^{0.137}R_{s,dw}^2-R_{\rm d}^2}.
\end{equation}

If either the dwarfs hot halo or disk experience ram pressure stripping the radius is reduced in proportion to the speed of shock induced \citep{Mori2000}
\begin{equation}
v_{\rm shock} = \frac{4}{3}v\sqrt{\frac{\rho_{\rm gal}}{\rho_{\rm halo/disk}}}.
\end{equation}
For simplicity we assume the gas is removed in shells and hence the change in radius due to this removal is half of $v_{\rm shock}$.
We also tested a different rate of ram pressure stripping $v_{\rm shock}\sim(2/3)c_{\rm s}$ from \citet{McCarthy2008} but found minimal variation in the final results.

The rate of change of the radius and hot halo can therefore be written as
\begin{eqnarray}
\dot{R}_{\rm h} &=& \dot{R}_{{\rm h},{\rm !RPS}} - 0.5v_{\rm shock}S_{\rm RPS,h},\\
\dot{R}_{\rm d} &=& -0.5v_{\rm shock}S_{\rm RPS,d}(R_{\rm d}>R_{\rm h}),
\end{eqnarray}
where all Boolean values are calculated as integers in the standard way, $1$ if true and $0$ if false.

\section{Results}

The model was run for varying virial masses of $M_{\rm vir}=3\times10^8$, $1\times10^{9}$, $3\times10^{9}$, $1\times10^{10}$, $3\times10^{10}$ and $1\times10^{11}$~$M_\odot$ in two halos of $z=0$ virial mass $1.37\times10^{12}$~$M_\odot$ \citep[consistent with the circular velocity of the RAVE survey;][]{Smith2007} and $2\times10^{12}$~$M_\odot$.
We note that under the chosen model of growth (where dwarfs do not undergo mergers or growth), the larger dwarfs may initially be more massive than the host galaxy (which does experience growth), we ignore any effects this may have on the model although note that it may impact the earliest redshifts.

\begin{figure*}
  \includegraphics[width=\textwidth]{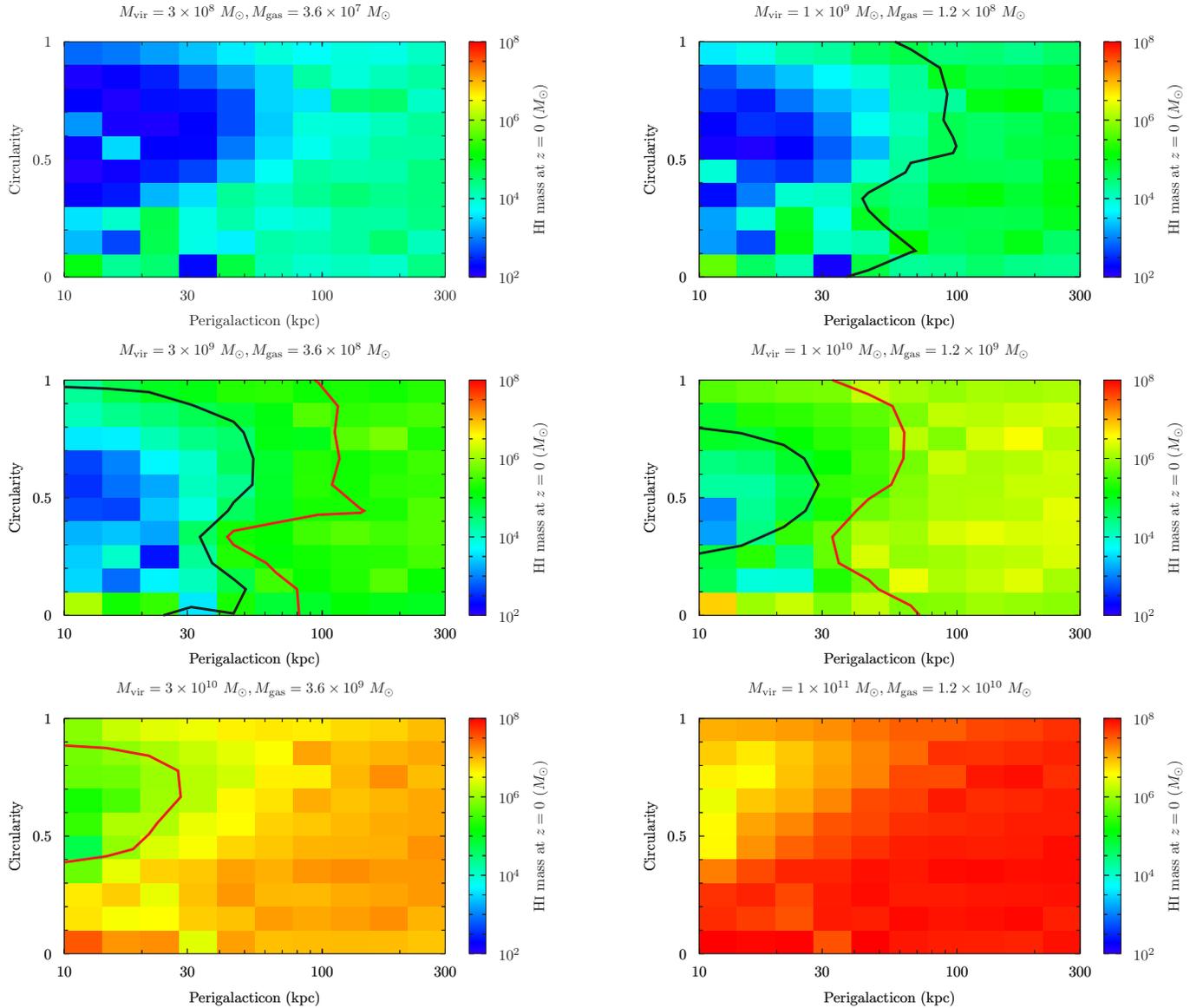}
  \caption{The present day H{\scriptsize I} mass for dwarfs around a host halo with $M_{\rm vir,0} = 1.37\times10^{12}~M_\odot$ \citep{Smith2007}, as a function of present day perigalacticon and circularity. Dwarfs of mass above $M_{\rm dw}=1\times10^{10}~M_\odot$ are not shown, as nearly all retain over $10^{5}~M_\odot$ of H{\scriptsize I}. The gas mass listed is the initial Hydrogen mass at $z=10$ and is a combination of H{\scriptsize I} and H{\scriptsize II}.  The color map is constructed using an inverse square interpolation from the randomly generated points. The star formation threshold is $N_{\rm HI}=10^{20}~$cm$^{-2}$. Contours of the smallest gas mass seen in the Local Group $(M_{\rm HI}=10^{5}~M_\odot)$ and the M81 group $(M_{\rm HI}=10^{6}~M_\odot)$ are shown as solid and dashed white (black and red solid) lines respectively. Unless closed, areas to the left are under the relevant gas mass. Even at masses below the SMC and LMC many halos will survive at low galactocentric radius and it is therefore unsurprising that they are as of yet unstripped. (A colored version of this figure is available online.)}\label{fig:dens}
\end{figure*}

Figure \ref{fig:dens} shows the effect of final perigalacticon and circularity of the dwarf on the amount of H{\small I} retained at the present day in a host halo of $M_{\rm vir,0}=1.37\times10^{12}~M_\odot$.
As may be intuitively expected, dwarfs with smaller perigalacticons lose a greater percentage of their initial mass.
Dwarfs with a medium circularity, ($\eta$$\sim$$0.5$), seem to lose more mass than those that with larger or smaller circularities.
Although such dwarfs spend less time in the denser part of the host halo than those on more circular orbits, the greater speed they obtain at perigalacticon may more than compensate for this.
There may be a similar effect at high perigalacticon, low circularity, where the extreme velocities encountered mean any tenuous medium is still able to strip some material from the dwarfs, however, there are very few dwarfs found in these regions and such an effect may just be noise.

\begin{figure*}
  \includegraphics[width=\textwidth]{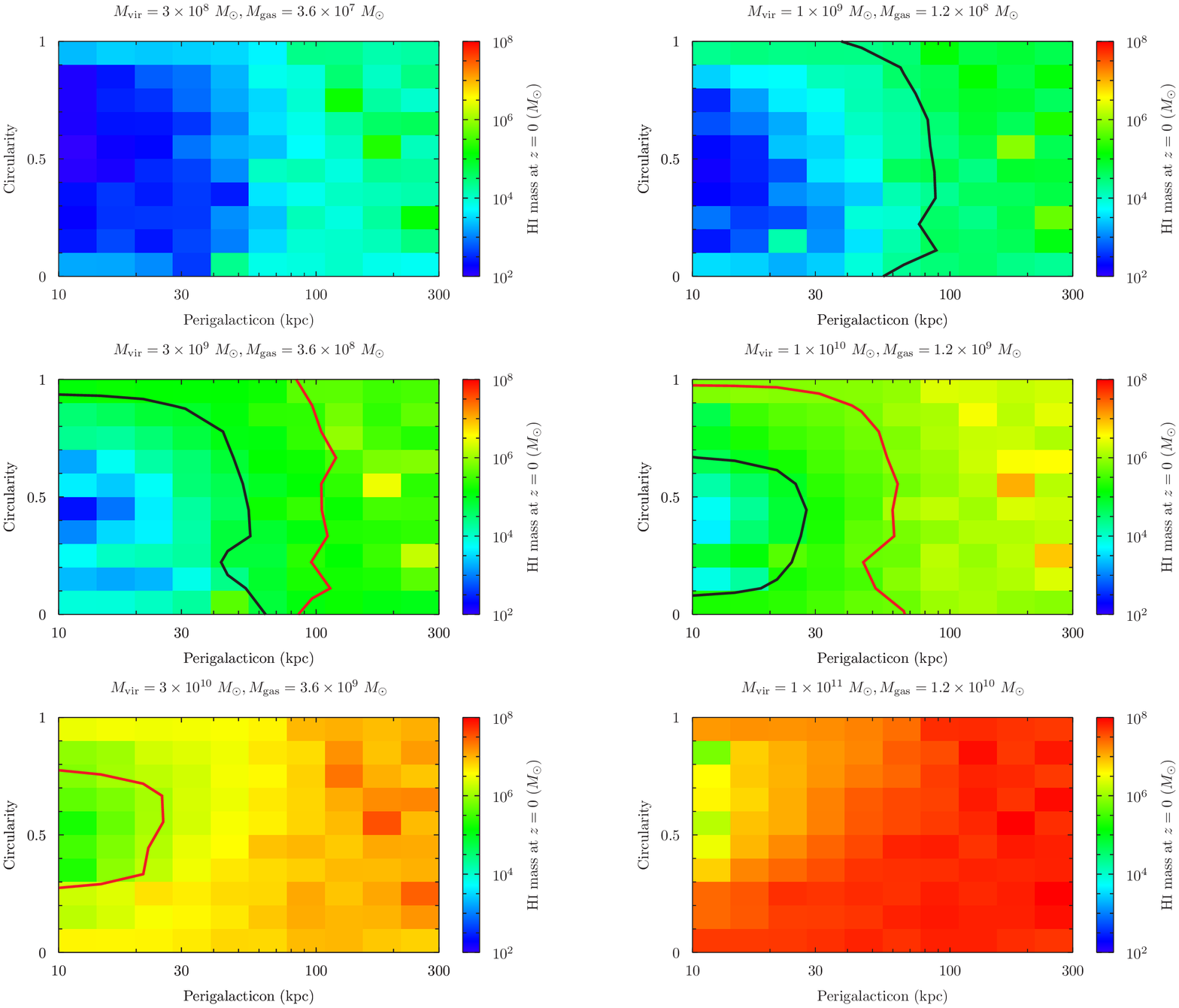}
  \caption{The present day H{\scriptsize I} mass for dwarfs around a host halo with $M_{\rm vir,0} = 2\times10^{12}~M_\odot$, as a function of present day perigalacticon and circularity. The gas mass listed is the initial Hydrogen mass at $z=10$ and is a combination of H{\scriptsize I} and H{\scriptsize II}.  The color map is constructed using an inverse square interpolation from the randomly generated points. The star formation threshold is $N_{\rm HI}=10^{20}~$cm$^{-2}$. Contours of the smallest gas mass seen in the Local Group $(M_{\rm HI}=10^{5}~M_\odot)$ and the M81 group $(M_{\rm HI}=10^{6}~M_\odot)$ are shown as solid and dashed white (black and red solid) lines respectively. Unless closed, areas to the left are under the relevant gas mass. (A color version of this figure is available online.)}\label{fig:dens2e12}
\end{figure*}

For dwarfs in a host halo $\sim$$50\%$ larger at $M_{\rm vir,0}=2\times10^{12}~M_\odot$ the result is much the same and is shown in figure \ref{fig:dens2e12}, with the only noticeable difference being a possible reduction in the gas mass for low circularity, low perigalacticon orbits.
Such a reduction likely occurs where the increased density, combined with the increased orbital speed in the more massive halo manages to strip these dwarfs where the previous host halo could not.

\subsection{Star Formation Threshold}\label{ssec:SFthresh}

Changing the star formation threshold dramatically changes the amount of gas that dwarfs are able to retain.
We examine this through a comparison of three threshold levels of H{\small I} required in the model for star formation ($10^{19}$~cm$^{-2}$, $10^{20}$~cm$^{-2}$ and $10^{21}$~cm$^{-2}$).

\begin{figure*}
\includegraphics[width=\textwidth]{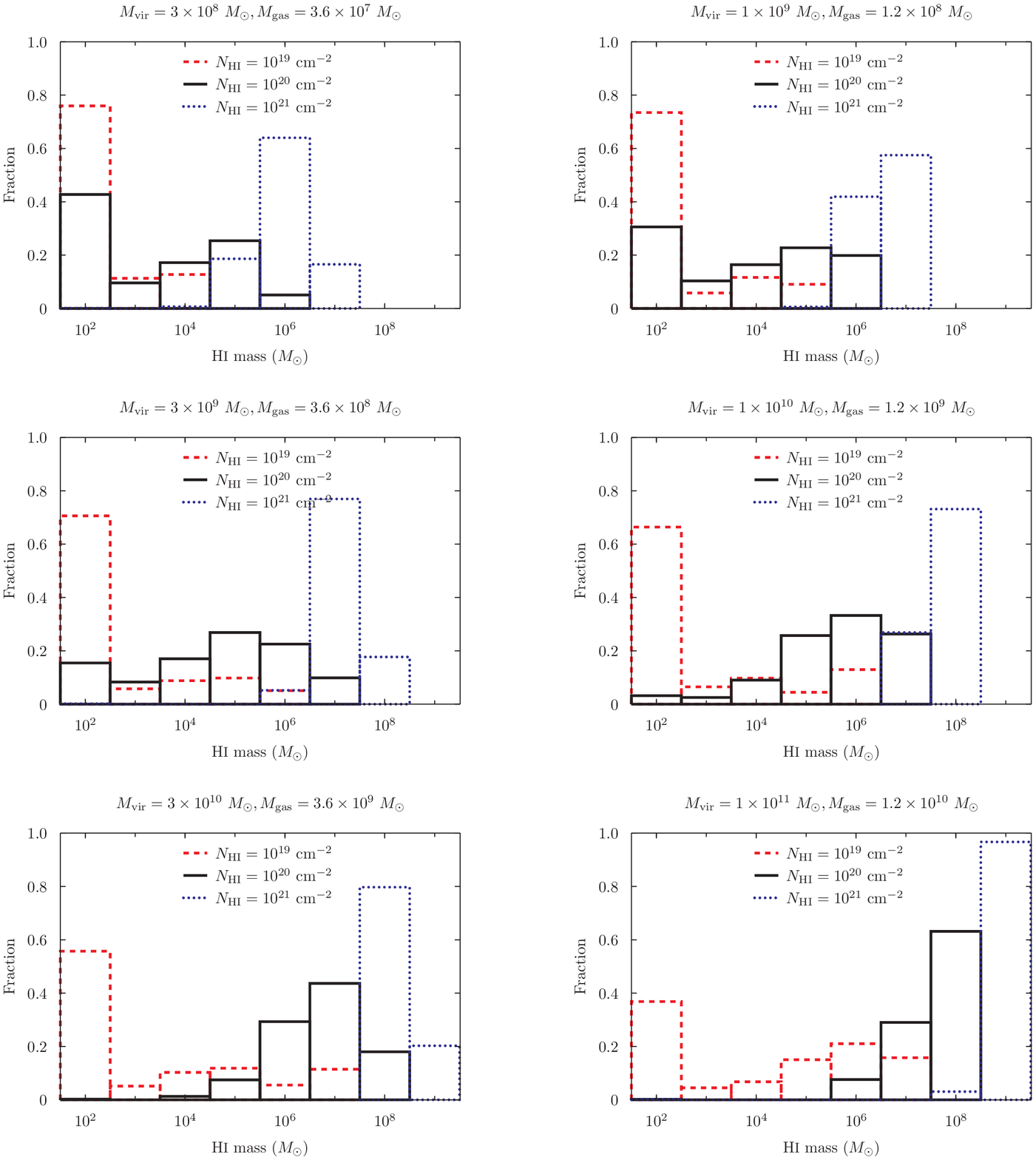}
\caption{Histogram of H{\scriptsize I} retained in dwarf galaxies with varying star formation thresholds. A cutoff of $10^{19}$~cm$^{-2}$ is shown as a (red) dashed line, $10^{20}$~cm$^{-2}$ as a black solid line and $10^{21}$~cm$^{-2}$ is shown as a (blue) dotted line. The amount of H{\scriptsize I} is displayed logarithmically with any gas mass below $10^{2}~M_\odot$ being included in that bin. The host halo was assumed to be a Milky Way like halo with a virial mass of $M_{\rm vir,0}=1.37\times10^{12}~M_\odot$. (A colored version of this figure is available online.)}\label{fig:SF}
\end{figure*}

Dwarfs with a low threshold ($N_{\rm HI}=10^{19}$~cm$^{-2}$) of star formation produce large amounts of star formation initially and subsequently blow out large amounts of gas through galactic winds.
Such an extended hot halo is then easily stripped by any passage through the host halo.
Dwarfs with a high threshold ($N_{\rm HI}=10^{21}$~cm$^{-2}$) experience the opposite problem.
Without strong winds forcing gas into a hot halo, the disk retains too much mass to be effectively stripped in all but the densest parts of the host halo.
The strength of these effects are shown in figure \ref{fig:SF}

\section{The Local Group and the M81 group}

\begin{figure*}
  \includegraphics[width=\textwidth]{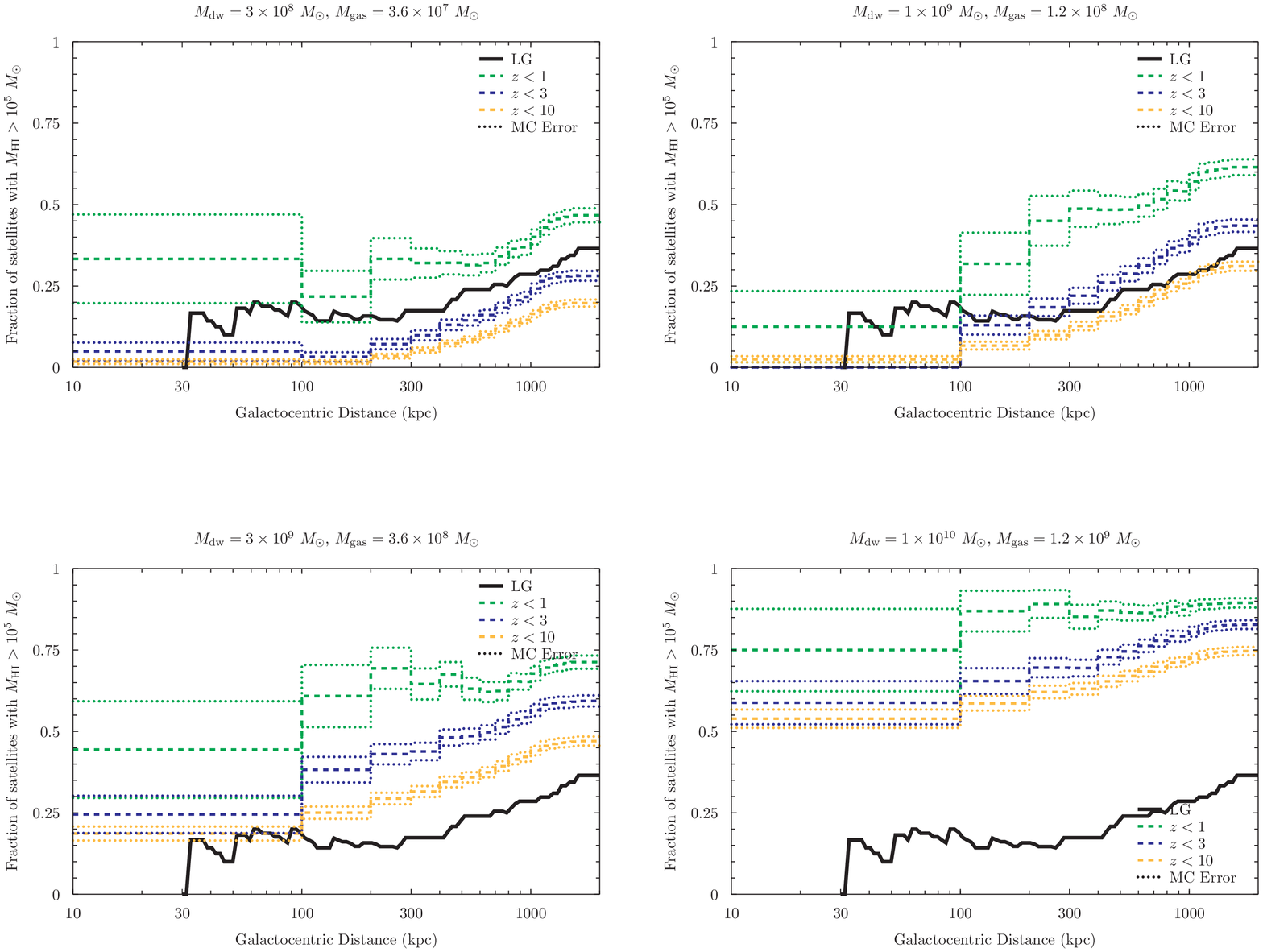}
  \caption{Fraction of dwarf galaxies below a radius that contains over $10^{5}~M_\odot$ of H{\scriptsize I} when orbiting a host galaxy of mass $M_{\rm vir,0}=1.37\times10^{12}~M_\odot$. The data for the Local Group is from \citet{Grcevich2009}. The accretion time is given by the time the dwarf first passed through the virial radius of the halo, with dwarfs that never do so being included in each. Dwarfs that accreted after redshift $1$ (or never) are shown with a dashed (green dashed) line, dwarfs which accreted after redshift $3$ (or never) are shown with a dot-dash (blue dashed) line and those which accreted after redshift $10$ (or never) are shown with a long-dashed (yellow dashed) line.
For clarity, we have binned the results into $100$~kpc bins with the resulting lines representing the end point of each bin (i.e. $100$--$200$~kpc gives the value at $200$~kpc).
In each case at high galactocentric distances dwarfs which accrete later have a higher fraction of gas that survives. At low galactocentric radius this is not necessarily true where low number statistics dominates. An estimate of the error given by the standard deviation of each dwarf from the line divided by the number of dwarfs is shown as dashed lines in the same color. (A colored version of this figure is available online.)}\label{fig:NBHplot}
\end{figure*}

We can use this model to obtain an estimate of the accretion time of the Local Group and other galaxy groups.
Most dwarf satellites within the Local Group are satellites of the Milky Way or M31.
Dwarfs around both galaxies are H{\small I} deficient within the virial radius of either \citep{Grcevich2009}.
We previously showed in NBH that such an effect implies that dwarfs begun accreting at an early time $(z=3$--$10)$.
In figure \ref{fig:NBHplot} we show the fraction of dwarfs that exceed $1\times10^{5}~M_\odot$ of H{\small I} at $z=0$ for the lowest mass dwarfs, defining the accretion time as the first time dwarfs passed through the virial radius of the host halo and including those that never pass through the virial radius in all cutoffs.
Compared to NBH, the accretion time suggested for the smallest mass dwarfs is more recent, a product of the strong dwarf winds that are now accounted for.

\begin{figure*}
  \includegraphics[width=\textwidth]{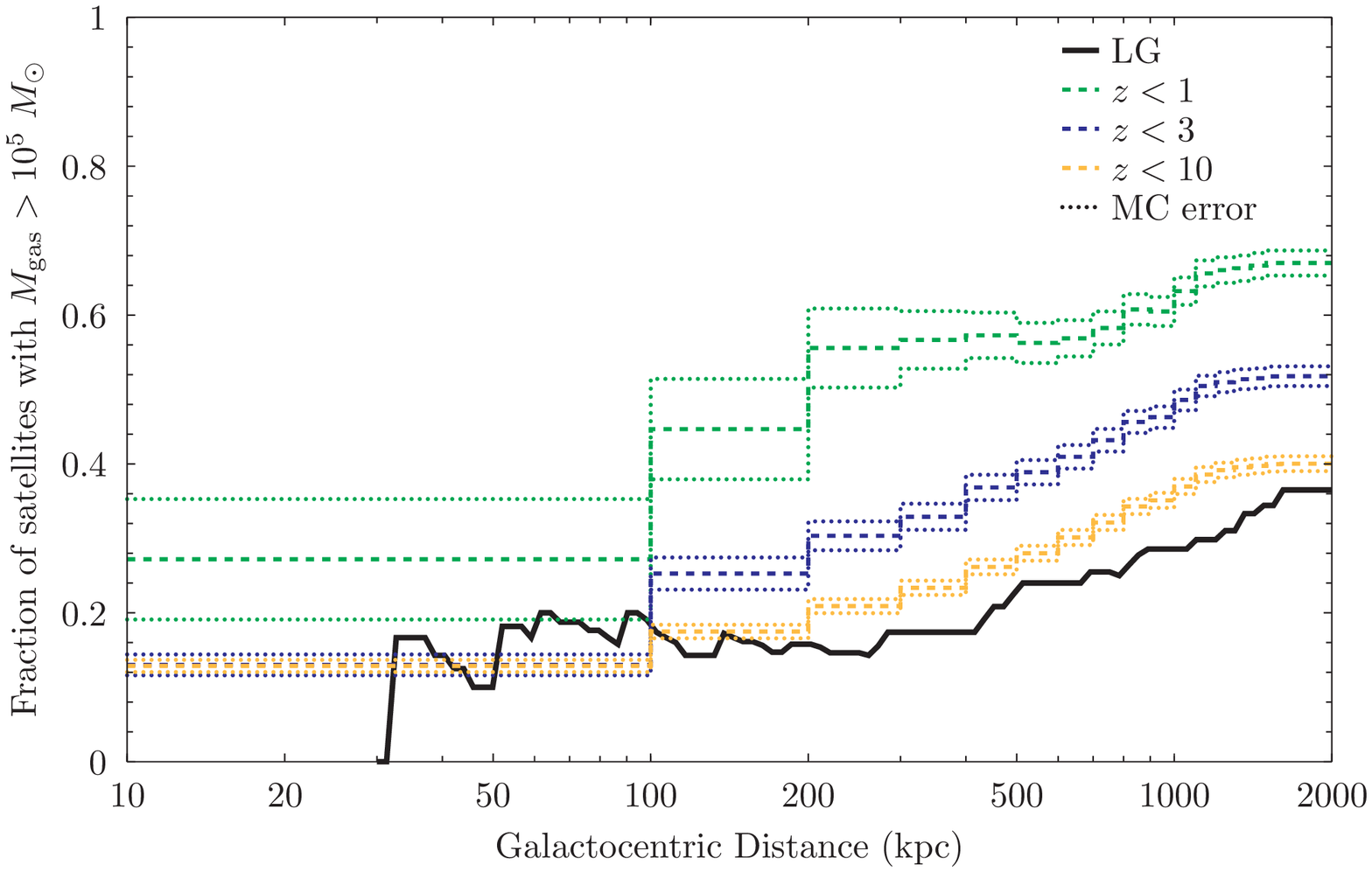}
  \caption{Fraction of dwarf galaxies below a radius that contain over $10^{5}~M_\odot$ of H{\scriptsize I} weighted according to the mass spectrum of dwarfs in Via Lactea II \citep{Diemand2007}. The redshift cutoffs are the same as figure \ref{fig:NBHplot}.}\label{fig:NBHmass}
\end{figure*}

Using the mass spectrum for dwarfs from Via Lactea II \citep{Diemand2007}  the expected fraction of all satellites can be reproduced and is shown in figure \ref{fig:NBHmass}.
Here, even accretion beginning at redshift of $10$ slightly overestimates the final fraction, a possible impact of the lower stellar mass of the model dwarfs compared to reality.
In addition, the Local Group is assumed to be observationally complete (or at least unbiased to mass/galactocentric distance).
Such a proposition is unlikely and it is probable that future detections of dwarfs will be the gas-deficient low mass objects as have recently been discovered. Such an overestimate therefore makes it more unlikely that the Milky Way begun accreting later than redshift $10$.

Detailed studies of external groups are only just beginning. To allow for a proper comparison,
we note that the M81 group survey is complete down to a limiting absolute magnitude of $M_{{\rm r}'}=-10$ \citep{Chiboucas2009}
and therefore choose this cut-off magnitude for the Local Group.
To determine the dwarf mass that satisfies this cut, we examine the integrated star formation over time and use the integrated stellar mass of Sculptor \citep[$M_{r}=-10.25$;][]{Jerjen1998} of $\sim$$1.2\times10^{6}$~$M_\odot$ \citep{Woo2008} to approximate this cut-off.
The closest model dwarf mass to this lifetime-integrated star formation is that of $M_{\rm dw}=3\times10^{9}$~$M_\odot$ and we therefore use it as the cut-off limit.
Such a mass is slightly above the common mass scale observed in the Local Group \citep{Strigari2008} with $M_{\rm dw}=3\times10^{9}$~$M_\odot$ corresponding to a mass of $4\times10^{7}$~$M_\odot$ within $300$~pc and this difference may reflect an inadequate star formation rate or be a consequence of baryonic processes lowering the central density (and therefore mass within $300$~pc).

\begin{figure*}
  \includegraphics[width=\textwidth]{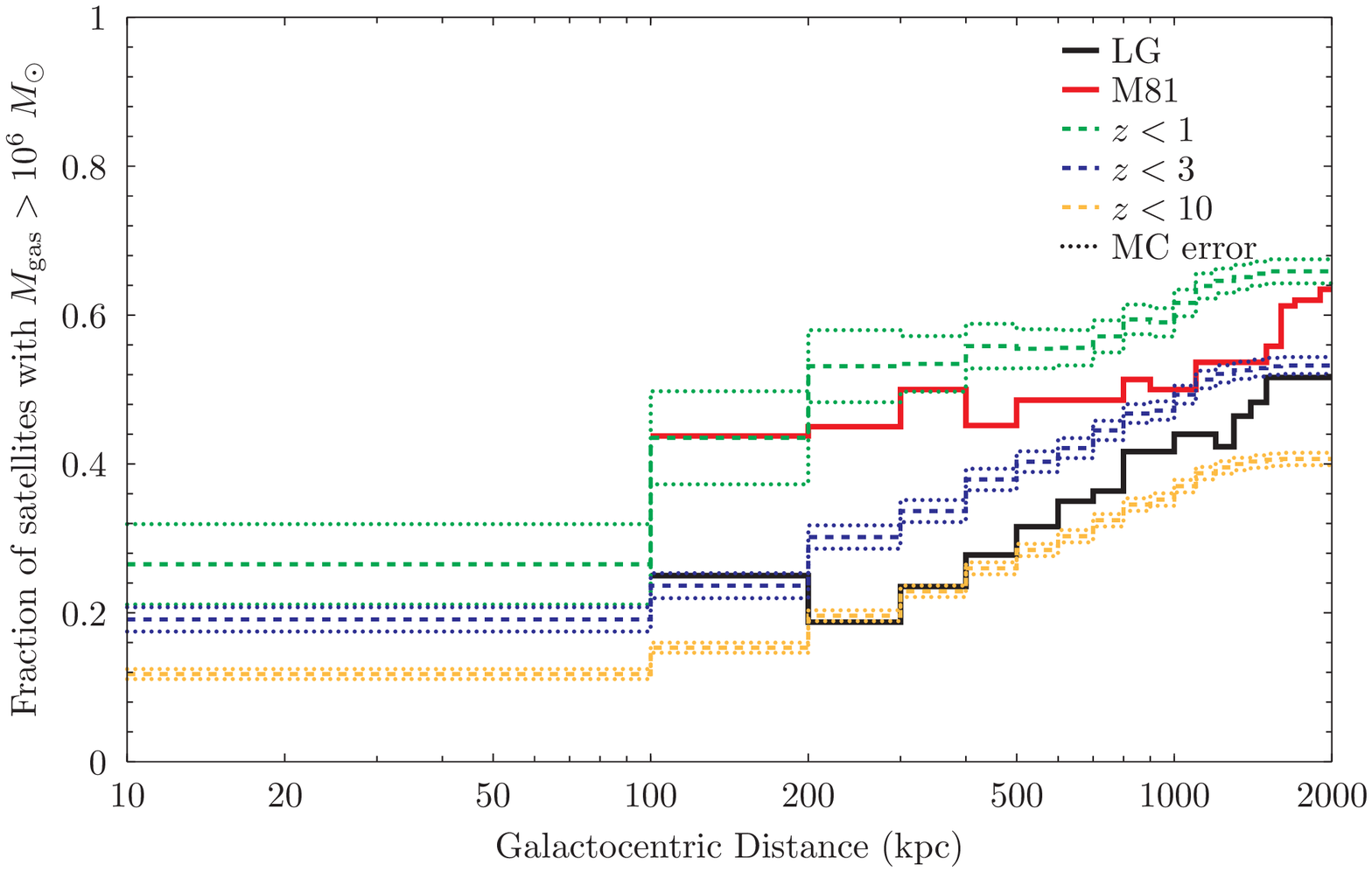}
  \caption{Fraction of dwarf galaxies below a given radius that contain over $10^{6}~M_\odot$ of H{\scriptsize I}. The fraction of observed galaxies that contain this in the Local Group is shown as a (black) solid line and the fraction in the M81 group shown as a (red solid) long-short dashed line. Only galaxies brighter than Sculptor are included. We bin these observed galaxies into $100$~kpc bins to reduce rapid changes in fractions due to the low number of galaxies involved. The model cut-offs are shown as dashed, dash-dot, long-dashed (dashed green, blue and yellow) lines as in figure \ref{fig:NBHplot}, with the binning of results the same. Below $\sim$$100$~kpc the groups are dominated by small number statistics, we subsequently leave this first bin off. (A color version of this figure is available online.)}\label{fig:LGM81}
\end{figure*}

We display the fraction of galaxies with $M_{\rm HI} > 10^{6}~M_\odot$ in the model against the M81 group and the Local Group in figure \ref{fig:LGM81}.
We use the H{\small I} masses from \citet{Karachentsev2007} and \citet{Roychowdhury2012}, and the distances from \citet{Karachentsev2002}.
The threshold is just above the lowest confirmed gas mass \citep[d1014+68 with $7.5\times10^{5}~M_\odot$ of H{\small I}][]{Roychowdhury2012} and we assume that dwarfs with only a published upper limit above this threshold fall below it.
We bin the results into $100$~kpc bins to smooth out the rapid changes that occur due to the small number of dwarfs involved.
Here the Local Group is satisfied by galaxies that begun accreting at redshift $z$$\sim$$3$--$10$, while the M81 Group is best fit by a later accretion epoch ($z=1$--$3$).
Within a few virial radii (where low number statistics may be more prevalant, but without any potential errors from association with the wrong host), M81 is best fit by accretion at approximately $z\sim1$ and the Local Group by $z\sim10$.

\begin{figure*}
  \includegraphics[width=\textwidth]{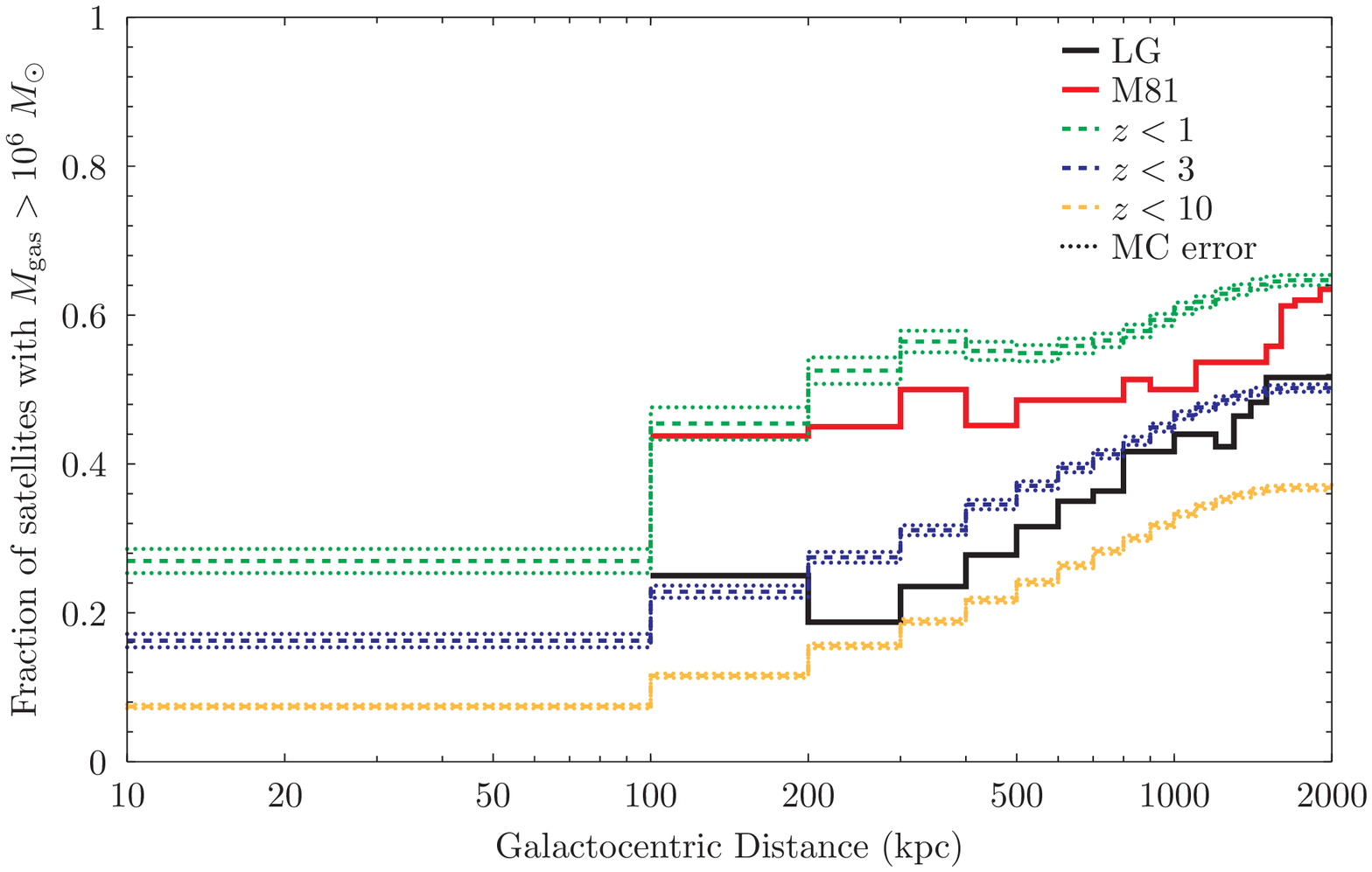}
  \caption{Fraction of dwarf galaxies below a given radius that contain over $10^{6}~M_\odot$ of H{\scriptsize I}. The fraction of observed galaxies that contain this in the Local Group is shown as a (black) solid line and the fraction in the M81 group shown as a long-short dashed (red solid) line. The model cut-offs are shown as dashed, dash-dot and long-dashed (dashed green, blue and yellow) lines as in fig \ref{fig:NBHplot}, with the binning of results the same. Below $\sim$$100$~kpc the groups are dominated by small number statistics, we subsequently leave this first bin off. The initial gas mass is set by $M_{\rm gas}=0.12~M_{\rm dw,acc}$ in contrast to figure \ref{fig:LGM81}. (A color version of this figure is available online.)}\label{fig:M81acc}
\end{figure*}

If the initial gas mass is limited by the dwarfs mass at $z=z_{\rm acc}$ ($M_{\rm gas}=0.12~M_{\rm dw,acc}$), the accretion time of the LG drops slightly, becoming closer to $z=3$ than $z=10$.
In this scenario, figure \ref{fig:M81acc}, both the Local Group and the M81 group require slightly more recent accretion.
The difference is slight however, as a large number of dwarf galaxies never pass the virial radius of the host galaxy.

\section{Conclusion}

The model presented is able to reproduce the observed fraction of gas-rich satellite dwarfs around a host halo.
Such a model is able to differentiate between accretion times between two galaxy groups and suggests that the satellites around the M81 group accreted at a later time.
Assuming that dwarf galaxies begun with a gas mass sufficient to give them the cosmic fraction of baryons to dark matter today if no gas had been lost suggests that the accretion of satellites by the largest galaxies in the Local Group (the Milky Way and M31) begun at an early redshift $z$$\sim$$10$.
Applying a magnitude cut-off to examine only the more massive dwarf galaxies suggests that they begun accreting at a similar time in the Local Group and that similar dwarfs inside the M81 group begun accreting more recently $z\sim 1-3$ and hence the M81 group started accreting its satellite galaxies $\sim2$~Gyr later than the Local Group.
Such a result may also be reflected in the masses of the HVC population with many large H{\small I} clouds \citep[with cloud masses up to $8\times10^{7}$~$M_{\odot}$;][]{Chynoweth2011} compared to the Local Group \citep[with a total HVC mass $\sim10^{8}$~$M_\odot$;][]{Lehner2012}.

Such a model although informative still has significant limitations when run on desktop machines.
Extending this model to Markov chain Monte Carlo algorithms is being investigated to remove or minimize some of these limitations.
The metallicity is assumed constant at a metallicity approximately that observed today in the most massive dwarfs \citep{Mateo1998}, such an assumption will lead to overcooling at the highest redshifts and in the smaller dwarfs and hence may be indicative of a later accretion time.
Acting in a similar way is the assumption of the initial gas mass, it is unlikely that the shallow potential well of a dwarf is able to gather such quantities of gas at high redshift, and the exact amount may vary on the accretion time of the dwarf.
The amount of gas initially should therefore be considered an upper limit, suggesting again a more recent accretion time than the model predicts.
The model also ignores the possibility of future low-redshift accretion or reaccretion of gas by dwarfs \citep[a distinct possibility, see][]{Ricotti2009,Nichols2012}, findings of significant gas accretion by satellites would be indicative of higher redshift accretion.
The dark matter potential of the dwarf is also assumed to be constant, a central density reduction in the dwarf will likely lead to gas being more easily removed from the centre and be indicative again of a more recent accretion time.
The gas distribution is assumed smooth, however, fractal distributions (more representative of the clumpy nature of dwarf disks) leads to faster stripping as instabilities build up which would require dwarfs to have been accreted more recently to produce the same distribution.
Furthermore, we have neglected tidal forces within this model.
As tidal forces decrease gas density at perigalacticon (where they are strongest) and as ram pressure stripping experiences a maximum at this same point, synergestic effects are possible.
Furthermore, we ignore some regions where tidal forces may be able to remove gas without the need for ram pressure stripping through synergestic effects between tides and supernova \citep{Nichols2013b}.
The accuracy of the analytic approximations will be compared with upcoming hydrodynamical simulations to allow for future calibration of this model.

Taken together it is likely that dwarfs accrete more recently than this model suggests but as these should happen in similar ways for all galaxy groups, such a model can be used to quickly determine an approximation of accretion time of a galaxy group and provides further evidence that even the local universe shows large differences in the accretion histories of nearby groups.

The dwarf model presented here may be refined further as proper motion measurements of dwarfs inside the Local Group improve, assisting the determination of accretion time \citep{Rocha2012}.
Such an improvement in proper motions seems unlikely for external groups.
Future multi-wavelength observations of the Local Group and external groups may also assist in the refinement of this model through narrowing of the allowed parameter space.
In particular, further observations of nearby groups \citep[e.g.][]{Chynoweth2011} will allow the extension of this model to more clusters, allowing a comprehensive study of the accretion of the nearby universe.

\acknowledgements

JBH is funded by a Federation Fellowship from the Australian Research Council.

\end{document}